\documentclass{iaus}
\usepackage{graphicx}
\usepackage{psfig}

\def\etal{et al.}

\def\teff{\ifmmode T_{\rm eff} \else $T_{\mathrm{eff}}$\fi}

\def\mh{$\mathrm{[M/H]}$}
\def\ltsima{$\buildrel<\over\sim$}
\def\lsim{\lower.5ex\hbox{\ltsima}}

\newcommand{\hii}{H~{\sc ii}}
\newcommand{\ha}{\ifmmode {\rm H}\alpha \else H$\alpha$\fi}
\newcommand{\hb}{\ifmmode {\rm H}\beta \else H$\beta$\fi}
\newcommand{\lya}{\ifmmode {\rm Ly}\alpha \else Ly$\alpha$\fi}

\newcommand{\heii}{He~{\sc ii}}
\newcommand{\Heiiuv}{He~{\sc ii} $\lambda$1640}

\newcommand{\la}{\raisebox{-0.5ex}{$\,\stackrel{<}{\scriptstyle\sim}\,$}}
\newcommand{\ga}{\raisebox{-0.5ex}{$\,\stackrel{>}{\scriptstyle\sim}\,$}}

\newcommand{\erg}{erg s$^{-1}$ cm$^{-2}$}
\newcommand{\parcminsq}{arcmin$^{-2}$}
\newcommand{\arcminsq}{arcmin$^{2}$}
\newcommand{\msunyr}{M$_\odot$ yr$^{-1}$}

\title[]
{Searching for Pop III stars and galaxies at high redshift}

\author[Daniel Schaerer]   
{Daniel Schaerer$^{1,2}$}

\affiliation{
$^1$ Geneva Observatory, University of Geneva,
51, Ch. des Maillettes, CH-1290 Versoix, Switzerland
\and
$^2$ Laboratoire d'Astrophysique de Toulouse-Tarbes, 
Universit\'e de Toulouse, CNRS,
14 Avenue E. Belin,
F-31400 Toulouse, France}

\pubyear{2008}
\volume{255}  
\pagerange{119--126}
\jname{Low-Metallicity Star Formation: From the First Stars to Dwarf Galaxies}
\editors{L.K. Hunt, S. Madden \& R. Schneider, eds.}
\begin{document}

\maketitle

\begin{abstract}
We review the expected properties of Pop III and very metal-poor starburst and the 
behaviour the \lya\ and \Heiiuv\ emission lines, which are most likely the best/easiest signatures
to single out such objects.
Existing claims of Pop III signatures in distant galaxies are critically examined, and the 
searches for \Heiiuv\  emission at high redshift are summarised.
Finally, we briefly summarise ongoing and future deep observations at $z > 6$
aiming in particular at detecting the sources of cosmic reionisation as well as
primeval/Pop III galaxies.
 
\end{abstract}

\firstsection 
              
\section{Introduction}
Finding the first stars and galaxies remains a major observational challenge in astrophysics.
Searches for metal-free (Pop III) or extremely metal-poor individual stars or for
ensembles of stars (clusters, proto-galaxies, populations of galaxies, etc.) are ongoing both
nearby (Pop III stars in the halo of our Galaxy), and in galaxies out to the highest
redshifts currently known. 

Astronomers have been quite inventive in searching for Pop III or metal-poor stars,
exploiting many possible direct and indirect signatures; some of these methods will be discussed
below. A more detailed account on primeval galaxies, some of the physics related to
these objects, an overview on searches etc.\ is presented in the lectures of 
Schaerer (2007).

Although some studies may have found signatures of Pop III stars, none of those
claims is very strong (see discussion below), and my personal opinion 
-- taking a somewhat conservative attitude -- is that Pop III still remain to be found.
However,  there is good hope that this goal should be reached in the quite near future
with current or forthcoming instrumentation, as I will sketch below.

\section{Expected properties of Pop III and very metal-poor populations}

\begin{figure}[tb]
\centerline{\mbox{\psfig{figure=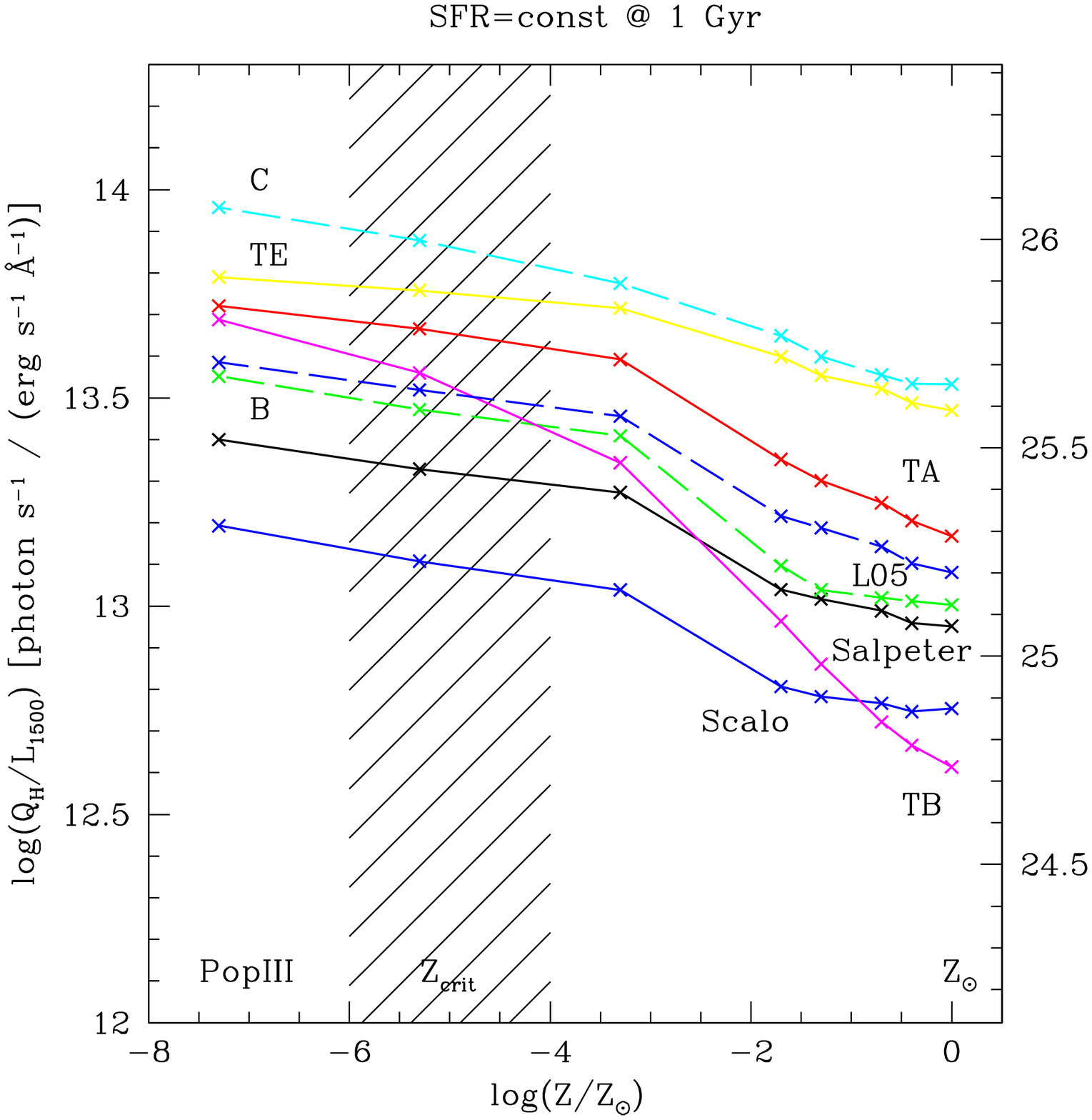,width=7.5cm}\hspace{0.5cm}
\psfig{figure=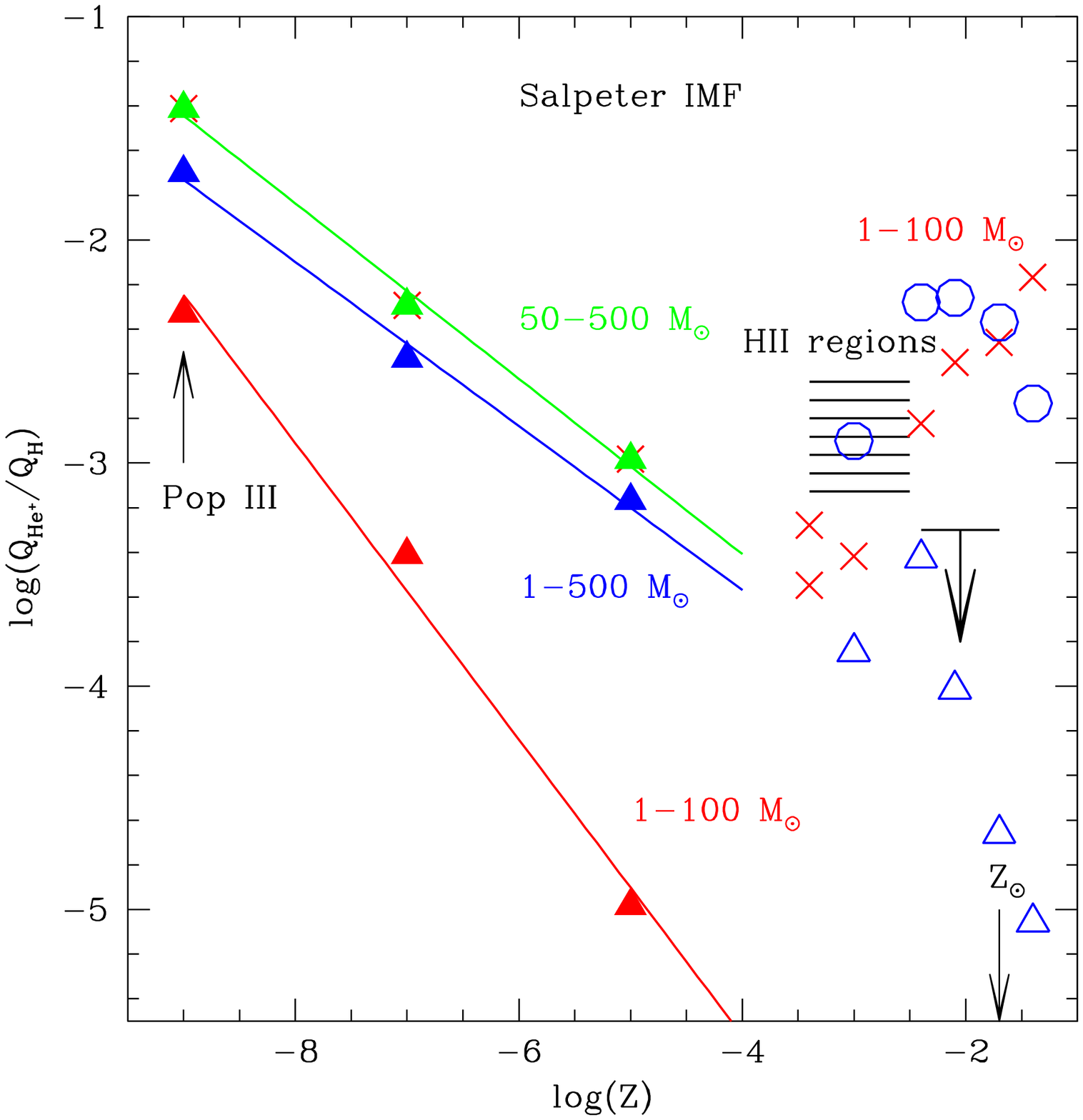,height=7.5cm}}}
\caption{{\bf Left:} Predicted ionising photon flux as a function of metallicity for different IMFs.
The ionising output is normalised here to the UV luminosity.
The shaded area labeled $Z_{\rm crit}$ indicates the domain where
the typical stellar mass / IMF may change (cf.\ Schneider et al.\ 2004).
{\bf Right:} Dependence of the hardness of the ionising flux, expressed by $Q_{\rm He^+}/Q_H$,
on metallicity and IMF. Figure from Schaerer (2003)}
\end{figure}
 
 Many authors have modeled the evolution of Pop III stars in the 1980s and before.
Since then the interest in these stars has been revived, and new generations of models
computed. Subsequently I will use predictions from the stellar evolution and non-LTE 
atmosphere models as well as  the evolutionary synthesis models computed by 
Schaerer (2002, 2003). For a detailed description
of the input physics and references to earlier work the reader is referred to these papers.
 
One of the main distinctive features of Pop III stars is their extreme compactness close to and
on the zero age main sequence (ZAMS), implying much higher effective temperatures (up to
$\sim$ 100 kK) than usual. For a given mass this leads to a considerable increase of the
ionising flux and to a much harder ionising spectrum (cf. Tumlinson et al.\ 2001, Schaerer 2002).
In addition, if the conditions in metal-poor environments favour the formation of massive stars
(cf. Bromm \& Larson 2004, and contributions by Abel, Bromm, Tan, and others in these
proceedings), the integrated spectrum of young, zero and very low metallicity stellar populations
can be quite different than that of populations at ``normal'' metallicities, as shown e.g. by
Schaerer (2002, 2003). In particular, the strong  \lya\ line emission and the presence
of nebular \heii\ emission lines are probably the best/easiest features to
search for very metal-poor and Pop III objects.

For example in Figure 1 (left) we show the increase of the ionising (Lyman continuum) photon flux
$Q_H$ normalised to the UV continuum luminosity with decreasing metallicity for different IMFs 
(Salpeter, Scalo; B, C as in Schaerer 2003, and other IMFs).
For a fixed IMF the increase from solar to zero metallicity is typically a factor of 2; assuming that
more extreme IMFs may be valid for Pop III, this increase can be up to a factor $\sim$ 10  (compared
to solar and Salpeter).
Note that this prediction also depends on the exact star-formation (SF) history, which affects
in particular the UV continuum output. 
The right panel of Fig.\ 1 shows the increase of the hardness of the ionising flux, expressed
by the ratio of He$^+$ to H ionising flux, $Q_{\rm He^+}/Q_H$, with decreasing metallicity.

As already mentioned, the strong Lyman continuum flux at low metallicities implies a strong
intrinsic \lya\ emission (cf. Schaerer 2002, 2003). For example, the {\em maximum \lya\ equivalent
width} predicted for an integrated stellar population increases from $\sim$ 200-300 \AA\
to 500-1000 \AA\ or even larger (depending on the IMF) from solar to zero metallicity, as shown 
in Fig.\ 2. Of course it must be reminded that after its emission inside the ionised region surrounding
the starburst \lya\ photons are scattered and may be absorbed, thereby altering the intrinsic
value of $W$ (see e.g.\ Schaerer 2007).

The increase of the hardness of the ionising radiation shown in Fig.\ 1 (right) translates into 
the increase of \heii\ recombination lines, such as \Heiiuv, shown in Fig.\ 2 (right). For obvious
reasons their strength depends on the exact SF history, age, and in particular on the IMF, as
shown by these figures (see also Schaerer 2003). While clearly a promising signature 
of very metal-poor populations as pointed out earlier\footnote{Caveat: \Heiiuv\ emission can also
of different origin (stellar cf.\ Sect. 3.1, AGN or shocks) and nebular \heii\ is observed in some 
normal \hii\ regions (see e.g.\  Thuan \& Izotov 2005) or in other peculiar objects (cf.\ Fosbury et al.\ 2003
and discussion in Schaerer 2003b).}, this strong sensitivity to the presence 
of the most massive stars -- i.e.\ to the IMF and age -- may complicate the interpretation of \heii\ lines, 
especially when dealing with non-detections (cf.\ below).

\begin{figure}[tb]
\centerline{\mbox{\psfig{figure=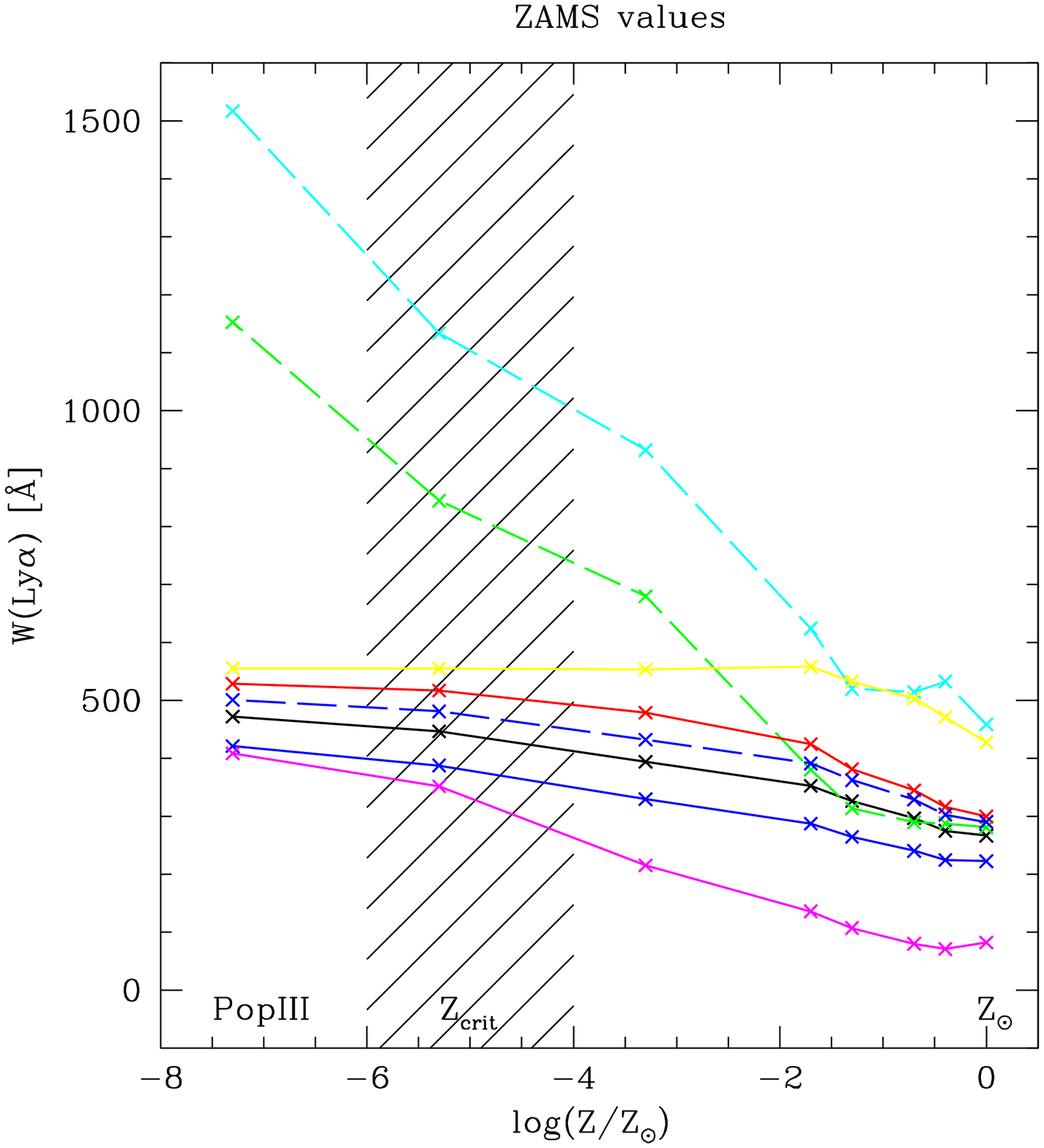,width=7.5cm}\hspace{0.5cm}
\psfig{figure=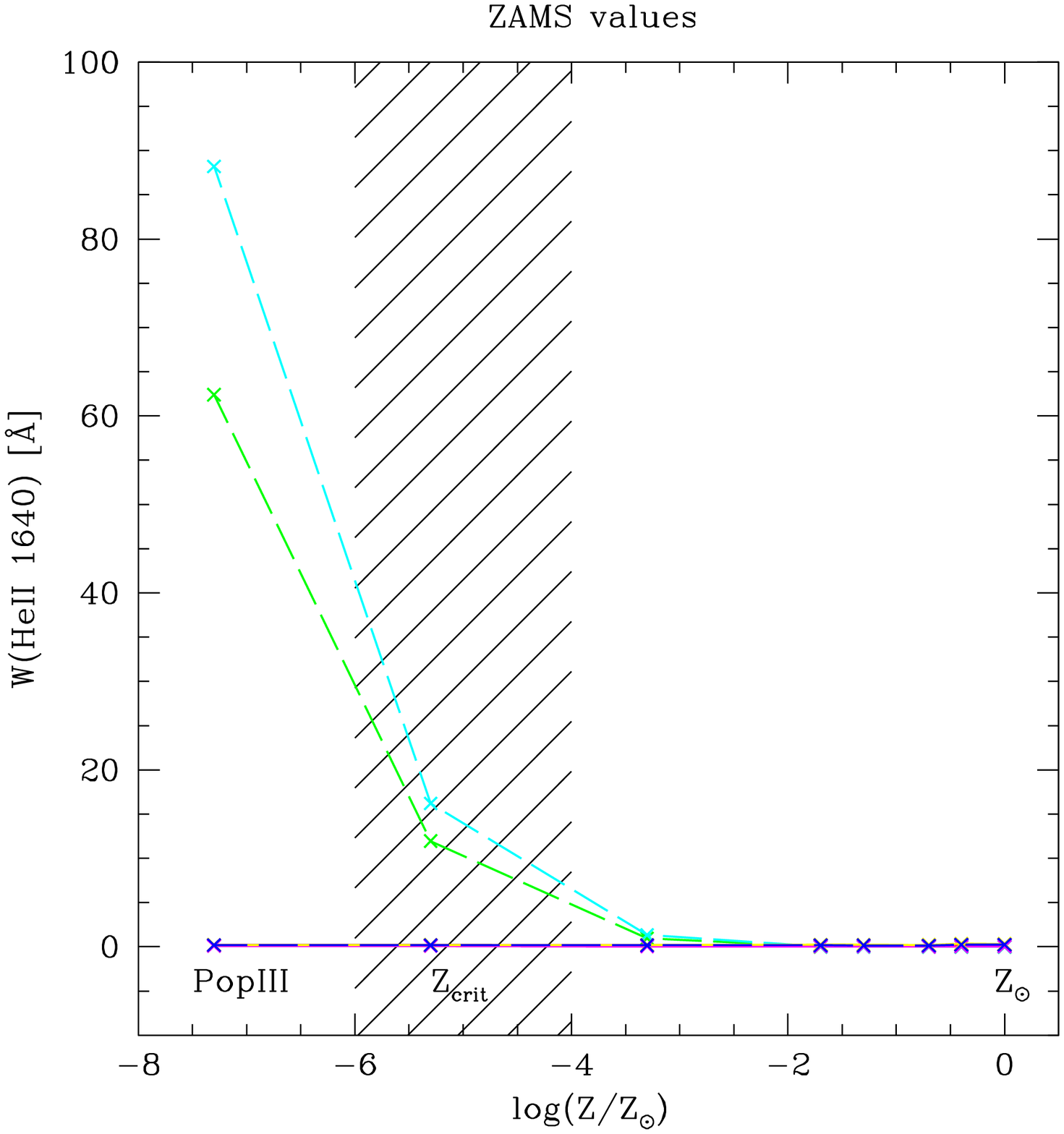,height=7.5cm}}}
\caption{{\bf Left:} Predicted maximum value of $W(\lya)$ as a function of metallicity and IMF (same colors
as Fig.\ 1).
{\bf Right:} Same as left for $W$(\Heiiuv$)$. See discussion in text.}
\end{figure}

Once the properties of individual Pop III clusters or galaxies are known, important questions
are of course how many such objects are expected and what is their distribution, e.g. in mass and 
luminosity? Answering these questions requires in particular a description of the transport and 
mixing of metals to determine the duration (locally) of Pop III SF events.
Several authors have addressed this using different approaches (e.g. Scannapieco et al.\ 2003,
Yoshida et al.\ 2004, Tornatore et al.\ 2007). The predicted Pop III SF rate, defined by SF
in regions with $Z < Z_{\rm crit}$, as a function of redshift from the recent computations
of Tornatore et al.\ (2007) is shown in Fig.\  3. Predicted number counts from Choudhury
\& Ferrara (2007) are shown in Fig.\ 5 for illustration.

\section{Have we found Pop III?}
It is possible that galaxies containing Pop III or very metal-poor stars have already been found
in different samples. Here I will critically examine some of these results.

\subsection{Have we found Pop III in Lyman Break Galaxies?}
For example, Jimenez \& Haiman (2006) have recently proposed an explanation to several
apparent puzzles concerning Lyman Break Galaxies (LBGs) and related objects at $z \sim$ 3--4.
They propose that their stellar populations contain $\sim$ 10--30 \% of primordial stars, which
would explain the observed  \Heiiuv\ emission in LBGs, 
the existence of some galaxies with very high W(\lya), 
the excess of Lyman continuum flux seen in some LBGs, and
the nature of \lya\ blobs.
 
 Although the idea to tackle simultaneously four problems is in principle attractive, 
 the proposed solution is probably not unique and not the most likely one, and some 
 of the 4 problems may not be robust observationally.
 First, the strength of the \Heiiuv\ line in the composite spectrum of LBGs is quite
 modest ($W(1640) =1.3 \pm 0.3$  \AA), and small enough to be explained by Wolf-Rayet
 stars in normal stellar populations (cf.\ models of Schaerer \& Vacca 1998, and
 Brinchmann et al.\ 2008). Furthermore the observed line is broad, as observed
 in Wolf-Rayet stars.
 Second, the existence of a population of galaxies at $z \sim$ 4--5 with high \lya\
 equivalent widths found by the LALA survey, does not seem to be confirmed by other
 groups or is at least very uncertain (see Sect.\ 3.2).
 Third,  other explanations are proposed for extended \lya\ blobs (e.g.\ Matsuda et al.\ 2007). 
Finally, the proposed explanation may also have difficulties with mixing time scales
(Pan \& Scalo 2007). 
The excess of Lyman continuum flux in some LBGs remains still puzzling
(see also Shapley et al.\ 2006, Iwata et al.\ 2008)

\subsection{Have we found Pop III in \lya\ emitters?}
A few years ago  Malhotra \& Rhoads (2002) found \lya\ emitters (LAE) at 
$z \sim 4.5$ with an unusual \lya\ equivalent width distribution from their LALA survey.
They suggested that their large fraction of objects with a high W(\lya) ($>$ 200-300 \AA)
could be due to very metal-poor or Pop III objects, a very unusual IMF, or to AGN.
Follow-up observations of these objects have been undertaken, also at X-rays,
and several papers have already adressed these results (see discussion in Schaerer 2007).
The AGN hypothesis has been rejected (Wang et al.\ 2004). Deep spectroscopy aimed at detecting
other emission lines, including the \Heiiuv\ line indicative of a
Pop III contribution (cf.\ Sect.\ 4), have been
unsuccessful (Dawson \etal\ 2004), although the achieved depth may not be sufficient.  
If taken at face value, the origin of these high W(\lya) remains thus unclear.  
However, there is some doubt on the reality of these
equivalent widths measured from NB and broad-band imaging, or at least
on them being so numerous even at $z=4.5$. 
First of all the objects
with the highest $W(\lya)$ have very large uncertainties since the
continuum is faint or non-detected. Second, the determination of
$W(\lya)$ from a NB and a centered broad-band filter ($R$-band in the
case of Malhotra \& Rhoads 2002) may be quite uncertain, e.g.\ due to
unknowns in the continuum shape, the presence of a strong spectral
break within the broad-band filter etc.\ (see Hayes \& Oestlin 2006
for a quantification, and Shimasaku \etal\ 2006 ).  Furthermore other
groups have not found such high $W$ objects (e.g.\ Hu \etal\ 2004, Ajiki
\etal\ 2003 and compilation in Verhamme et al.\ 2008) suggesting also 
that this may be related to insufficient
depth of the LALA photometry.

Recently Shimasaku \etal\ (2006) and Dijkstra \& Wyithe (2007) have again,
used $W(\lya)$ distributions to argue for a non-negligible contribution
of Pop III stars in LAEs, this time in objects at $z \ge 5.7$.
In fact the {\em observed} restframe values of $W(\lya)$  LAE samples at 
$z=5.7$ and 6.5 obtained with SUBARU are considerably lower than 
those of Malhotra \& Rhoads at $z=4.5$, and only few objects 
show $W^{\rm rest}_{\rm obs}(\lya) \ga 200$ \AA, the approximate
limit expected for ``normal'' stellar populations (cf.\ Fig.\ 2).
However, if the IGM transmission $T_\alpha$ affects half of the \lya\ line,
as often assumed (see also Hu et al.\ 2004), this would imply true intrinsic 
equivalent widths $2/T_\alpha$ times higher than the observed value, 
i.e. typically 4 times higher at $z=5.7$!
For this reason Shimasaku \etal\  suggest that these may be young 
galaxies or objects with Pop III contribution. Based on this reasoning and on
modeling the \lya\ luminosity function and IGM transmission, Dijkstra \& Wyithe (2007) 
argue for the presence of $\sim$ 4--10 \% of Pop III SF in $\sim$ half of \lya\ 
selected galaxies and in few percent of i-drop galaxies at $z \ga$ 5.7.

Although possible, this conclusion rests strongly on the assumption
that the IGM transmission (due to individual or overlapping \lya\ forest clouds) really
cuts out a significant fraction of the \lya\ line emerging from the galaxy.
This is by no means clear, e.g. since it implies the need for cold gas close
(and/or infalling) to the galaxies, which has not been observed yet. E.g.\
in the LBG cB58 studied in depth, Savaglio et al.\ (2002) find no \lya\ absorption
close to the galaxy (at $\Delta v \la$ 2000 km s$^{-1}$). 
Also Verhamme et al.\ (2008)
find no need for IGM absorption in their \lya\ profile modeling of LBG/LAE at $z \sim$
3--5. Furthermore, absorbing
a fraction of \lya\ by the IGM is even more difficult if the intrinsic
\lya\ profile emerging from the galaxy is already redshifted (and asymmetric),
as expected if outflows  are as ubiquitous  as in LBGs at lower redshift.
Assessing these important issues remains to be done. In the meantime conclusions
depending strongly on corrections to the observed \lya\ equivalent widths (or LF)
should probably be taken with caution.

\begin{figure}[tb]
\centerline{\psfig{figure=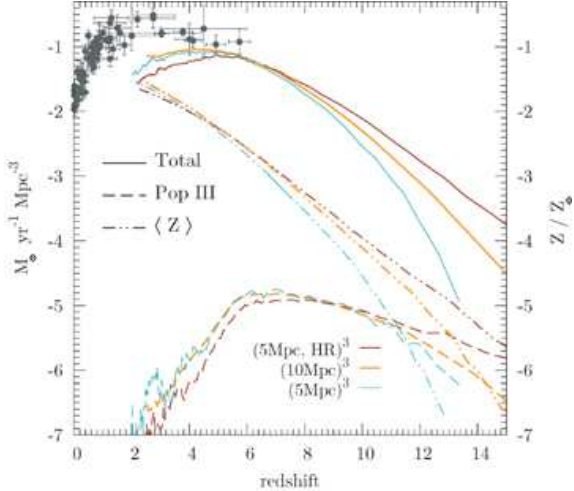,height=6.5cm}}
\caption{Predited SFR density and average metallicity as a function of redshift 
showing the contribution of Pop III (defined as metallicity $Z < Z_{\rm crit}$, cf.\ above)
and other stars. 
Recently Nagao et al.\ (2008) have derived $SFRD_{\rm Pop III} \protect\la 5.\, 10^{-6}$ \msunyr\ Mpc$^{-3}$ at 
$4.0\protect \la z \protect\la 4.6$ close to the predicted value at this redshift.
Figure from Tornatore et al.\  (2007)}
\end{figure}

\section{Searches for \heii\ emission}
Several groups have tried to use the expected \Heiiuv\ emission (cf.\ Sect. 2) to search for Pop III stars,
so far with no positive detection. However, the limits obtained from most recent survey 
begin to provide interesting constraints.

\subsection{Upper limits from individual objects or composite spectra} 
Follow-up spectroscopy of LALA sources by Dawson et al.\ (2004)
(cf.\ Sect.\ 3.2) have yielded an upper limit
of $W(\lya) <$ 25 \AA\ (3 $\sigma$) and \Heiiuv/\lya$<$ 0.13--0.20 at 2--3 $\sigma$ from their 
composite  spectrum (11 objects).
Nagao \etal\ (2005) searched for \Heiiuv\ with deep spectroscopy of one strong \lya\ emitter at 
$z=6.33$. Currently the lowest  limit on \heii\ emission is that measured by Ouchi et al.\
(2008) in composite spectra of 36 and 25 $z=3.1$ and 3.7 LAEs. They reach 
$I($\Heiiuv$)/I($\lya$) <$ 0.02 and 0.06 ($3 \sigma$) at these two redshifts, which translates to 
$\log(Q_{\rm He^+}/Q_H) < -1.9$ and $<-1.4$ respectively\footnote{The approximate translation
between line ratios and ionising flux ratios is $Q_{\rm He^+}/Q_H =  0.6 \,\, I($\Heiiuv$)/I($\lya$)$
(cf.\ S03).}. Although no definite conclusions can be drawn from that, comparison with Fig.\ 1 (right) 
show that these limits are already close
to the maximum or slightly below the predicted values for metal-free populations with
IMFs including very massive stars.

\begin{figure}[htb]
\centerline{\psfig{figure=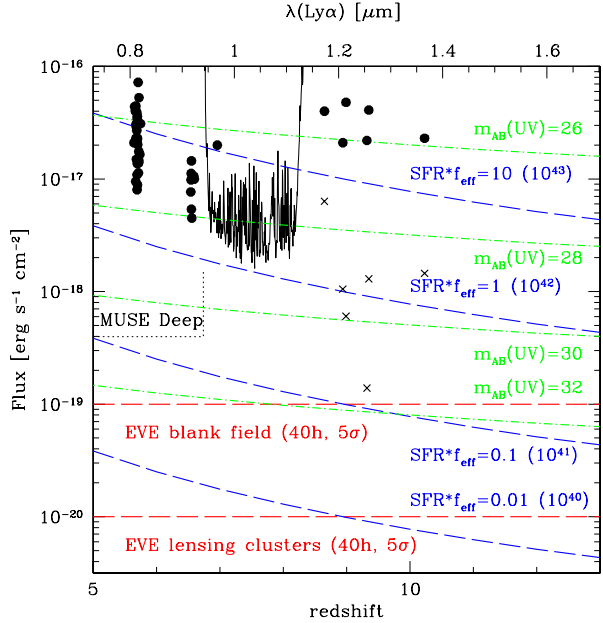,height=8.5cm}}
\caption{Observed \lya\ flux as a function of redshift (observed wavelength on top) achieved 
with 10m class telescopes.
Observations from Shimasaku et al.\ (2006), Kashikawa et al.\ (2006), and Ota et al.\ (2008)
at $z=5.7$, 6.5, and 7, as well as the $z\sim$ 8.5 to 10 lensed candidates of Stark 
et al.\ (2007) are shown by the filled circles (observed fluxes, and crosses: intrinsic
fluxes for lensed objects). The black line shows the flux limit from deep 
NIRSPEC/Keck spectroscopy of Richard et al.\ (2008) from 0.95 to 1.1 $\mu$m.
Blue dashed lines show the fluxes corresponding to values of 
SFR$\times f_{\rm eff}$ from 0.01 to 10 \msunyr, where $f_{\rm eff}$ denotes the effective \lya\ transmission, 
including the IGM transmission and other possible losses, e.g.\ inside the galaxy.
Green dash-dotted lines show the expected \lya\ flux for star-forming galaxies
with rest-frame UV continuum magnitudes from 26 to 32 in the AB system, assuming 
$f_{\rm eff}=1$.
In both cases the conversion to \lya\ flux was computed assuming a standard
SFR calibration based on Kennicutt (1998) and case B.
Deep observations with MUSE, a 2nd generation instrument for the VLT, and observations with
EVE, a proposed multi-object spectrograph for the E-ELT, will allow to push the current
limits down by $\sim$ 1--2 magnitudes!}
\end{figure}

\subsection{The first constraint on the \Heiiuv\ flux density at high redshift}
Using an original approach combining narrow/intermediate-band filters to select dual
\lya\ + \Heiiuv\ emitters at $z \sim 4.0$ and 4.6, Nagao et al.\ (2008) have carried out a 
survey of 875 arcmin$^2$ using SUBARU in the Subaru Deep Field
(see also Nagao, these proceedings, for more details).
No such dual emitters were found, down to a flux limit of $(6-7) \times 10^{-18}$ \erg\
for \Heiiuv. With some assumptions on the IMF and using the models of S03, the observed flux 
limits translate to an upper limit for the Pop III SFR of $\ga 2$ \msunyr, and a SFR
density of $SFRD_{\rm Pop III} \la 5.\, 10^{-6}$ \msunyr\ Mpc$^{-3}$ at $4.0 \la z \la 4.6$ 
(Nagao et al. 2008). Interestingly this first Pop III SFR density determination turns out
to be very close to the theoretical  prediction of Tornatore et al. (2007) shown in Fig.\ 3.
Clearly further work and deeper observations are required to track the elusive Population
III at high-redshift and future theoretical work may provide a more detailed/accurate
picture of the expectations.

\section{Ongoing and future deep observations at $z>6$}

The impressive depth -- down to few times $10^{-18}$ \erg\ -- reached already in emission 
line searches with SUBARU, VLT, and Keck
is illustrated in Fig.\ 4, where \lya\ measurements and upper limits
obtained in the visible and near-IR domain are compiled. In the case of lensed galaxies
the intrinsic depths can be considerably larger, as e.g.\ shown for the candidate
high-z galaxies of Stark et al.\ (2007).
In several cases of $z>6$ galaxies or candidates, spectroscopic follow-up
observations have already been undertaken to search for various emission lines,
including the potential Pop III indicator \Heiiuv\ discussed above (e.g. Pell\'o et al.\ 2005, Stark et al.\ 2007, 
Richard et al.\ 2008, in preparation).

Since the average metallicity of galaxies must decrease with increasing redshift, searches
for primeval galaxies and/or Pop III galaxies are also naturally focussed towards the
highest redshift. At the same time galaxy searches at $z>6$ are also of great interest
to identify the sources of cosmic reionisation.

 \begin{figure}[htb]
\centerline{\psfig{figure=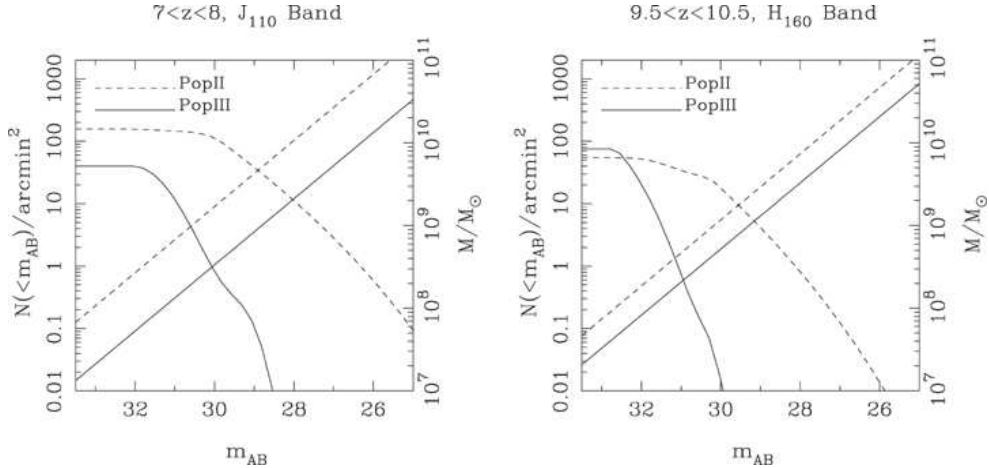,width=13cm}} 
\caption{Predicted (cumulative) number density of "normal" (Pop II) and Pop III dominated sources
for $z \sim 7.5$ (left) and $z \sim 10$ (right). Figure from Choudhury \& Ferrara (2007).}
\end{figure}

Reviewing the status of galaxy searches/observations at $z>6$ is beyond the scope 
of the present contribution. For an overview see e.g.\ Schaerer (2007 and references therein).
The most recent results from searches for $z>7$ galaxies are presented by
Bouwens et al.\ (2008) and Richard et al.\ (2008), who use surveys in blank fields
or fields with massive galaxy clusters benefiting from strong gravitational lensing.
At the bright end of the LF the number density of galaxies at $z\ga7$, and hence also the total
SFR density, remains controversial (cf.\ Richard et al.\ 2006, Bouwens et al.\ 2008).
At fainter magnitudes ($m_{AB} \ga$ 27) the results from these two
approaches (blank fields and lensing clusters) give consistent number densities 
(see Richard et al.\ 2008). The lensing cluster technique allows one to extend the
observations to the faintest levels, currently down to an effective 
magnitude of $m_{AB} \sim$ 29--30. 
Clearly the two techniques,
one with an additional gain in depth by lensing over a small area and the other over wide
fields, are complementary and both approaches will likely play an important role 
in finding distant, primeval, maybe even Pop III dominated objects.
 
The lensing studies show a source density of 3--100 \parcminsq $(\Delta z = 1)^{-1}$ for $7 < z < 8$ 
from $m_{AB} = 28$ to 30 (Richard et al.\ 2008). 
This source density is also in agreement with the theoretical predictions 
from Stiavelli et al.\ (2004) and Choudhury \& Ferrara (2007), shown in Fig.\ 5 . 
If spectroscopy of objects down to magnitudes of 30 or even 32 become feasible
with extremely large telescopes (cf.\ Fig.\ 4) this means that we should expect very high source 
densities, e.g.\ several hundred objects(!) in a field of view of several \arcminsq\
as foreseen for ELT instruments. High multiplex multi-object spectrographs reaching 
with an optical and near-IR coverage up to $\sim 1.7 \mu$m, such
as EVE proposed for the E-ELT, should therefore be very efficient in detecting
\lya\ and \Heiiuv\ lines up to redshift 13 and 9.4 respectively.
Great progress can be expected from searches for the sources of cosmic reionisation and 
primeval/Pop III galaxies in the fairly near future!



\begin{thebibliography}{}
\bibitem[]{} Ajiki, M., et al., 2003, \textit{AJ}, 126, 2091
\bibitem[Bouwens et al.(2008)]{2008arXiv0803.0548B} Bouwens, R.~J., 
Illingworth, G.~D., Franx, M., 
\& Ford, H.\ 2008, 
arXiv:0803.0548 
\bibitem[Brinchmann et al.(2008)]{2008MNRAS.385..769B} Brinchmann, J., 
Pettini, M., \& Charlot, S.\ 2008, \textit{MNRAS}, 385, 769 

\bibitem[Bromm 
\& Larson(2004)]{2004ARA&A..42...79B} Bromm, V., \& Larson, R.~B.\ 2004, \textit{ARAA}, 42, 79 

\bibitem[Choudhury 
\& Ferrara(2007)]{2007MNRAS.380L...6C} Choudhury, T.~R., \& Ferrara, A.\ 2007, \textit{MNRAS}, 380, L6 
\bibitem[]{} Dawson, S., et al.\ 
2004, \textit{ApJ}, 617, 707 
\bibitem[Dijkstra 
\& Wyithe(2007)]{2007MNRAS.379.1589D} Dijkstra, M., \& Wyithe, J.~S.~B.\ 2007, \textit{MNRAS}, 379, 1589 
\bibitem[Fosbury et al.(2003)]{2003ApJ...596..797F} Fosbury, R.~A.~E., et 
al.\ 2003, \textit{ApJ}, 596, 797 
\bibitem[]{} Hayes, M., 
\"Ostlin, G.\ 2006, \textit{A\&A}, 460, 681 
\bibitem[]{} Hu, E.~M., et al. \ 2004, \textit{AJ}, 127, 563
\bibitem[]{}  Iwata, I., et al., 2008, ApJL, submitted, arXiv:0805.4012
\bibitem[]{} Jimenez, R., \& Haiman, Z.\ 2006, Nature, 440, 501 
\bibitem[]{}  Kashikawa, N., et al. 2006, ApJ, 648, 7
\bibitem[]{}  Kennicutt, R.C., 1998, \textit{ARAA}, 36, 189
\bibitem[]{} Malhotra, S., Rhoads, J.E., 2002, \textit{ApJ}, 565, L71
\bibitem[Matsuda et al.(2007)]{2007ApJ...667..667M} Matsuda, Y.,  et al.,
\& Petitpas, G.~R.\ 
2007, \textit{ApJ}, 667, 667 
\bibitem[Nagao et al.(2008)]{2008ApJ...680..100N} Nagao, T., et al.\ 2008, 
\textit{ApJ}, 680, 100 
\bibitem[Ouchi et al.(2008)]{2008ApJS..176..301O} Ouchi, M., et al.\ 2008, 
\textit{ApJ}s, 176, 301 
\bibitem[Pan 
\& Scalo(2007)]{2007ApJ...654L..29P} Pan, L., \& Scalo, J.\ 2007, \textit{ApJ}, 654, L29 
\bibitem[Pell{\'o} et al.(2005)]{2005IAUS..225..373P} Pell{\'o}, R., 
Schaerer, D., Richard, J., Le Borgne, J.-F., 
\& Kneib, J.-P.\ 2005, in "Gravitational Lensing Impact on Cosmology", IAU Symp.\ 225, 373 
\bibitem[Richard et al.(2008)]{2008arXiv0803.4391R} Richard, J., Stark, 
D.~P., Ellis,  R.~S., et al.
\& Smith, G.~P.\ 2008, \textit{ApJ}, in press, arXiv:0803.4391 

\bibitem[Savaglio et al.(2002)]{2002ApJ...567..702S} Savaglio, S., Panagia, 
N., \& Padovani, P.\ 2002, \textit{ApJ}, 567, 702 

\bibitem[]{} Scannapieco, E., Schneider, R., \& Ferrara, A.\ 2003, \textit{ApJ}, 589, 35 
\bibitem[]{} Schaerer, D. 2002, \textit{A\&A}, 382, 28
\bibitem[]{} Schaerer, D. 2003, \textit{A\&A}, 397, 527
\bibitem[]{} Schaerer, D. 2003b, in "Multi-wavelength cosmology" , astro-ph/0309528
\bibitem[]{} Schaerer, D., 2007, in "The emission line Universe" , XVIII Canary Islands
Winter School of Astrophysics, Ed.\ J. Cepa, Cambridge Univ.\ Press, arXiv.0706.0139
\bibitem[Schaerer 
\& Vacca(1998)]{1998ApJ...497..618S} Schaerer, D., \& Vacca, W.~D.\ 1998, \textit{ApJ}, 497, 618 
\bibitem[]{} Schneider, R., Ferrara, A., \& Salvaterra, R.\ 2004, \textit{MNRAS}, 351, 1379 
\bibitem[Shapley et al.(2006)]{2006ApJ...651..688S} Shapley, A.~E., 
Steidel, C.~C., Pettini, M., Adelberger, K.~L., 
\& Erb, D.~K.\ 2006, \textit{ApJ}, 651, 688 
\bibitem[]{} Shimasaku, K., et 
al.\ 2006, \textit{PASJ}, 58, 313 

\bibitem[]{} Stark, D.P., Ellis, R.S., Richard, J., Kneib, J.-P., Smith, G.P., Santos, M.R.,
2007, \textit{ApJ}, 663, 10
\bibitem[Stiavelli et al.(2004)]{2004ApJ...600..508S} Stiavelli, M., Fall, 
S.~M., \& Panagia, N.\ 2004, \textit{ApJ}, 600, 508 
\bibitem[Thuan 
\& Izotov(2005)]{2005ApJS..161..240T} Thuan, T.~X., \& Izotov, Y.~I.\ 2005, \textit{ApJS}, 161, 240 
\bibitem[Tornatore et al.(2007)]{2007MNRAS.382..945T} Tornatore, L., 
Ferrara, A., \& Schneider, R.\ 2007, \textit{MNRAS}, 382, 945 
\bibitem[]{} Tumlinson, J., Giroux, M.L., Shull, J.M., 2001, \textit{ApJ}, 550, L1

\bibitem[Verhamme et al.(2008)]{2008arXiv0805.3601V} Verhamme, A., 
Schaerer, D., Atek, H., \& Tapken, C.\ 2008, \textit{A\&A}, in press, arXiv:0805.3601 
\bibitem[]{} Wang, J.~X., et al.\ 2004, \textit{ApJ}, 608, L21 
\bibitem[Yoshida et al.(2004)]{2004ApJ...605..579Y} Yoshida, N., Bromm, V., 
\& Hernquist, L.\ 2004, \textit{ApJ}, 605, 579 

\end{thebibliography}
\end{document}